%% file: main.tex
\documentclass{article}

\PassOptionsToPackage{numbers, compress}{natbib}

\usepackage[final]{neurips_2019}

\usepackage[utf8]{inputenc} 
\usepackage[T1]{fontenc}    
\usepackage{hyperref}       
\usepackage{url}            
\usepackage{booktabs}       
\usepackage{amsfonts}       
\usepackage{nicefrac}       
\usepackage{microtype}      
\usepackage{bm}
\usepackage[pdftex]{graphicx}
\usepackage{amsmath}
\usepackage{amsthm}
\newtheorem{theorem}{Theorem}[section]
\title{BPMR: Bayesian Probabilistic Multivariate Ranking}

\author{%
  Nan Wang\\
  Department of Computer Science\\
  University of Virginia\\
  Charlottesville, VA, 22903 \\
  \texttt{nw6a@virginia.edu} \\
  \And
  Hongning Wang\\
  Department of Computer Science\\
  University of Virginia\\
  Charlottesville, VA, 22903 \\
  \texttt{hw5x@virginia.edu} \\  
}

\begin{document}

\maketitle

\begin{abstract}

Multi-aspect user preferences are attracting wider attention in recommender systems, as they enable more detailed understanding of users' evaluations of items. Previous studies show that incorporating multi-aspect preferences can greatly improve the performance and explainability of recommendation. However, as recommendation is essentially a ranking problem, there is no principled solution for ranking multiple aspects collectively to enhance the recommendation. 

In this work, we derive a multi-aspect ranking criterion. To maintain the dependency among different aspects, we propose to use a vectorized representation of multi-aspect ratings and develop a probabilistic multivariate tensor factorization framework (PMTF). The framework naturally leads to a probabilistic multi-aspect ranking criterion, which generalizes the single-aspect ranking to a multivariate fashion. Experiment results on a large multi-aspect review rating dataset confirmed the effectiveness of our solution. 

\end{abstract}

\input{intro}
\input{method}
\input{exp}
\input{conclusion}


\bibliographystyle{abbrv}
\bibliography{reference.bib}

\clearpage
\input{appendix}

\end{document}

%% file: intro.tex
\section{Introduction}
To understand users' detailed preferences beyond merely their overall evaluation of an item, multi-aspect preferences are extensively collected and studied for recommendation. For example, users' detailed ratings on value, location, and service aspects of a hotel. Previous studies show that incorporating such multi-aspect preferences in addition to overall preference can greatly boost the performance of rating prediction, and also promote the explainability of the recommendation \cite{viewpoint, EFM, Diao:2014:JMA:2623330.2623758, multiverse, bptf, HongImproving,tao2019fact,yang2021saer}. However, as recommendation is essentially a ranking problem \cite{Sun2017MRLRMR,MTER,wang2020direct}, 
how to rank items with respect to multiple aspects collectively remains unsolved. Existing ranking criteria, such as Bayesian personalized ranking (BPR) \cite{BPR,Sun2017MRLRMR}, are only designed for single-aspect ranking. Therefore, they can only be applied to overall rating or to each aspect separately. We believe that multi-aspect preferences should be organically integrated for improved ranking performance, where the key challenge is to enable multi-aspect comparison and maintain their dependence.

On the other hand, tensor factorization (TF) based methods have shown its natural fitness and effectiveness in modeling multi-aspect data \cite{TensorDecomp, MTER, multiverse, tagRec}. We seek solution from TF for multi-aspect preference ranking. The idea behind TF-based models is that each slice of a tensor corresponds to the preference of users over a particular aspect. The properties of users, items and aspects can be represented by latent factor matrices of $U\in \mathbb{R}_+^{M\times d^U}$, $V\in \mathbb{R}_+^{N\times d^V}$, $W\in \mathbb{R}_+^{K\times d^W}$ respectively, where $M,\,N,\,K$ are the numbers of users, items and aspects, and $d$s are the dimensions of latent factors. The preference prediction under each aspect in the factorized tensor could then be measured by the distance of latent factors defined by the specific factorization method, as $X_{uil} =<U, V, W>_{uil}$. 
To optimize for recommendation, the objective of factorization ought to maximize the probability of observed ranking orders among items. 
But there is no principled solution for ranking multiple aspects collectively; and in existing solutions the ranking is simply performed according to the predicted overall rating \cite{MTER}, or to each aspect separately \cite{tagRec}. 
Moreover, almost all previous TF-based solutions model users' preferences over multiple aspects as independent, by defining the likelihood of observing multi-aspect preferences as a point-wise product of each aspect \cite{PMF, Rai:2015:SPT:2832747.2832775, bptf, temporal}. As a result, the dependence among multiple aspects is ignored in ranking.


In order to compare across multi-aspect preferences collectively, it is necessary to take a vectorized representation of all aspect preferences from a user to an item and view the vector as an integrated observation of preference. This new perspective makes it possible to model the intrinsic dependence among aspects, and such dependence also enables aspect-based explanations of recommendations. 
We first formulate a vectorized multi-aspect ranking criterion that enables the modeling of multi-aspects in the ranking task. To capture the dependence among aspects, we propose a probabilistic tensor factorization framework (PMTF) for modeling the generation of aspect ratings as sampling from a multivariate Gaussian distribution:
\begin{equation*}
    \bm R_{ui} \sim \mathcal{N}(<U,V,W>_{ui}, \Sigma_{ui}),\,\,\,\, \bm R_{ui}\in \mathbb{R}^K
\end{equation*}
The correlation among aspects is explicitly introduced in PMTF by the covariance matrix $\Sigma_{ui}$, and will enable multi-aspect preference ranking. Based on this proposed framework, we derive a Bayesian probabilistic multivariate ranking solution (BPMR), which models the comparison between observed pair-wise orders of vectors in a multivariate fashion. We directly maximize the likelihood of the observed pair-wise orders between vectors to optimize the ranking of all aspects collectively. From an information-theoretic perspective \cite{l2x, Kullback1959Information}, we show that by explicitly modeling the correlation, our method provides an information-theoretic regularization for the consistency and explainability of multi-aspect preference modeling and ranking.

In Section \ref{sec:MR}, we formulate the multi-aspect ranking criterion by extending the single-aspect ranking criterion.  We then propose the PMTF framework which leads to the derivation of the multi-aspect ranking solution BPMR in Section \ref{sec:bpmr}. A specialized Expectation Maximization algorithm is proposed to learn both the latent factors and covariance matrices. We show in Section \ref{sec:exp} that both quantitative and qualitative experimental comparisons against several state-of-the-art methods on a large TripAdvisor multi-aspect rating dataset demonstrate the effectiveness of our solution. 

%% file: method.tex
\section{Muti-aspect Ranking Criterion}
\label{sec:MR}
Suppose we have $M$ users, $N$ items, $K$ aspects, and user preferences on those aspects are indicated by explicit ratings. We use $P$ to denote the set of all users and $Q$ for the set of all items. Each observation is a vector of aspect ratings from user $u$ to item $i$, denoted as $R_{ui} = (r_{ui1}, r_{ui2}, ..., r_{uiK})$. 
For simplicity, we use CP Decomposition \cite{TensorDecomp} as a specific tensor factorization method to illustrate our framework; but it can be generalized to other tensor factorization method, e.g., Tucker decomposition. Three matrices $U\in\mathbb R_+^{M\times d}$, $V\in\mathbb R_+^{N\times d}$, $W\in\mathbb R_+^{K\times d}$ are used to represent latent factors of users, items and aspects, respectively, with each row vector representing a single entity accordingly. The predicted multi-aspect ratings are then obtained as $\hat R_{ui} = (U_u * V_i)W^T$, where $*$ denotes the element-wise (Hadamard) product. 

\subsection{Reformulation of Single-aspect Ranking}
The foundation of single-aspect ranking criterion is to maximize the probability of observing ground-truth pair-wise orders specified by the overall preference \cite{BPR,wang2021prs}. To formalize it, we need to first construct an order set $D_S: P\times Q\times Q$ by $D_S := \{(u,i,j)|i >_u j\}$. The meaning of $(u,i,j) \in D_S$ is that user $u$ prefers item $i$ over item $j$. We assume that observed items in a user's rating history ($r>0$) are preferred over unobserved items ($r=0$). The order of each pair could be treated as a Bernoulli variable, i.e., $p(\bm r_{ui}>\bm r_{uj}) +p(\bm r_{ui}\leq\bm r_{uj}) = 1$. Then the likelihood of observed pair-wise orders is formulated as:
\begin{equation}
\label{eq:BPR}
    \prod\limits_{u\in P}p(>_u|\Theta) = \prod\limits_{(u,i,j)\in P\times Q\times Q} p(\bm r_{ui}>\bm r_{uj}|\Theta)^{\delta((u,i,j)\in D_S)}
    \cdot p(\bm r_{ui}>\bm r_{uj}|\Theta)^{\delta((u,i,j)\notin D_S)}
\end{equation}
where $\delta(b)$ is an indicator function which is 1 if $b$ is true and 0 otherwise. $\bm r_{ui}$ is the random variable from which the observed rating $r_{ui}$ of user $u$ to item $i$ is sampled. However, it is not straightforward to define an order set for vectors of multi-aspect observations. To extend it to a multi-aspect ranking criterion, we first rewrite the likelihood for single aspect ranking and get rid of $D_S$ by expressing the observed pair-wise order as the sign of the difference between two observed preferences:
\begin{equation}
    \prod\limits_{u\in P}p(>_u|\Theta) = \prod\limits_{(u,i,j)\in P\times Q\times Q}p\big((\bm r_{ui} -\bm r_{uj})(r_{ui} -r_{uj})>0|\Theta\big)
\label{eq:BPR-reform-nosign}
\end{equation}
Intuitively, we require the predicted user preferences over a pair of items to be consistent with the observed preferences. By properly specify the underlying distribution of $\bm r_{ui}$, different weights are induced to the ordered pairs by the observed preference difference. 

\subsection{Vectorized Multi-aspect Ranking Criterion}
For multi-aspect explicit preferences, each observation is denoted as a vector: $R_{ui} = (r_{ui1},\dots,r_{uik})$ from user $u$ to item $i$. A straightforward way is to apply Eq. \eqref{eq:BPR-reform-nosign} on each aspect separately and then maximize the product of resulting probability across all aspects. However, such a solution implicitly assumes a user's evaluation of different aspects is independent from each other. This is clearly not true as numerous studies have proved the dependency among different rating aspects in users' opinionated evaluation of items \cite{wang2010latent,MTER}. Instead, we take an integrated view of vector comparison, to capture such dependency. We extend Eq. \eqref{eq:BPR-reform-nosign} to multi-aspect explicit preferences collectively by,
\begin{align}
\label{eq:multi-bpr}
    \prod\limits_{u\in P}p(>_u|\Theta) &= \prod\limits_{(u,i,j)\in P\times Q\times Q}p(\sum\limits_{l=1}^K(\bm r_{uil} - \bm r_{ujl}) (r_{uil} - r_{ujl})>0|\Theta) \\
                &= \prod\limits_{(u,i,j)\in P\times Q\times Q}p\big((\bm{R}_{ui} -\bm{R}_{uj})\cdot ({R}_{ui} -{R}_{uj})>0|\Theta\big)\nonumber
\end{align}
Here we assume all aspects are equally important for item ranking. But if it requires different emphasis on different aspects, we can apply a weighting vector $\bm w\in \mathbb R^K$ and take the Hadamard product as: $p\big((\bm{R}_{ui} -\bm{R}_{uj})\cdot(({R}_{ui} -{R}_{uj})*\bm w)>0\big)$. To simplify our notations, we denote $\bm D_{uij} = \bm R_{ui} -\bm R_{uj}$ as a random vector from which the observed difference $D_{uij} = R_{ui} - R_{uj}$ is generated. Then the ordering of two observed rating vectors is defined by the difference vector $D_{uij}$. This formulation of multi-aspect ranking leads to a natural geometric interpretation of vector comparison: maximizing the dot product between $\bm D_{uij}$ and $D_{uij}$ corresponds to requiring the samples generated from $\bm D_{uij}$ to a direction aligned with the observed vector $D_{uij}$. In other words, it requires  the difference between the predicted rating vectors to be consistent with the observed difference in the multi-aspect preferences. 

Direct optimization of Eq. \eqref{eq:multi-bpr} is not feasible without explicitly defining the probability $p(\bm D_{uij}\cdot D_{uij}|\Theta)$. Most previous work proposes to use a link function to map the prediction from a vector to a scalar, and then optimize the resulting objective function \cite{BPR,Sun2017MRLRMR}. This unfortunately cannot capture the dependence among those multiple rating aspects. In the next section, we show that by taking a probabilistic view of tensor factorization for multi-aspect rating modeling, we can directly quantify  the ranking criterion in a probabilistic sense, with the consideration of dependence among aspects.

\section{Bayesian Probabilistic Multivariate Ranking}
\label{sec:bpmr}
To capture the dependence among different aspects, we assume the aspect ratings are sampled from a multivariate Gaussian distribution. Based on the factorized multi-aspect tensor, the observed multi-aspect rating vector from user $u$ to item $i$ could be viewed as being generated from:
\begin{equation}
    \bm R_{ui} \sim \mathcal{N}((U_u * V_i)W^T, \bm\Sigma_{ui}),\,\,\,\, \bm R_{ui}\in \mathbb{R}^K
\end{equation}
Intuitively, the emphasis of different users on each aspect may vary, and the same logic applies to items. Therefore, the covariance matrix should be personalized for users and items to capture their detailed preferences.
To formally quantify the personalized covariance composition for the observation of a user-item pair, we take a perspective stemmed from the Item Response Theory (IRT) \cite{Fayers2004Item,Reid2007Modern}. IRT is originally from psychometrics, which states that the probability of a specific
response is determined by the item’s attribute and the subject’s individuality \cite{Lin}. Mapping back to our problem, we assume that the covariance matrix of an observed rating vector is a result from two independent channels of preference emphasis from the corresponding user and item. Specifically, we assign each user $i$ and each item $j$ a personalized covariance matrix $\bm\Sigma_i^U$ and $\bm\Sigma_j^V$; then the covariance matrix of a pair of user-item is given by:
\begin{equation}
    \bm\Sigma_{ui} = \lambda\bm\Sigma_u^U + (1-\lambda)\bm\Sigma_i^V
\end{equation}
where $\lambda \in [0,1]$ is a hyper-parameter, reflecting the relative importance of user covariance in this composition. Then the conditional distribution over the observed aspect rating vectors is specified as,
\begin{equation}
    p(R|U,V,W,\bm\Sigma) = \prod\limits_{u=1}^M\prod\limits_{i=1}^N\big[\mathcal{N}\big(R_{ui}|(U_u * V_i)W^T, \lambda\bm\Sigma_u^U + (1-\lambda)\bm\Sigma_i^V\big)\big]^{I_{ui}}
\label{eq:pmtf}
\end{equation}
where $\mathcal{N}(\bm{x}|\bm{\mu}, \bm\Sigma)$ is the probability density function of the multivariate Guassian distribution with mean $\bm \mu$ and covariance $\bm\Sigma$. Note that there might be missing aspect ratings in $R_{ui}$, and we can apply a binary indicator vector $\bm I_{ui}\in \mathbb{R}^K$ to the difference between the observed and predicted aspect rating vector in Eq. \eqref{eq:pmtf} as $\big(R_{ui} - (U_u*V_i)W^T\big)*\bm I_{ui}$. Compared with existing probabilistic matrix/tensor factorization methods \cite{PMF}, the covariance matrix in Eq. \eqref{eq:pmtf} is no longer a hyper-parameter to set, but important statistics to be learnt from the observations. Although we use the multivariate Gaussian distribution for modeling the dependence among aspects, other kinds of multivariate distributions, such as multivariate Poisson distribution, could also be applied to characterize the generation of multi-aspect rating vectors. This leads to a general probabilistic multivariate tensor factorization framework (PMTF), which takes the dependence among aspects into account. 

\subsection{Multi-aspect Ranking with PMTF}
In previous works, the probability defined on the difference between user preferences is often obtained by a link function, such as the sigmoid \cite{BPR} or softmax \cite{Sun2017MRLRMR}. However, with PMTF, we can directly get the probability distribution of $\bm D_{uij}$ by assuming that different items are independently evaluated by a user:
\begin{equation}
    p(\bm D_{uij} = D_{uij}|U,V,W,\bm\Sigma) = \mathcal{N}\big(D_{uij}|(U_u * (V_i-V_j))W^T, \bm\Sigma_{uij}\big) 
\end{equation} 
where $\bm\Sigma_{uij} = \bm\Sigma_{ui} + \bm\Sigma_{uj}$. By taking the dot product of $\bm D_{uij}$ and $ D_{uij}$, it naturally leads to the explicit expression of Eq. \eqref{eq:multi-bpr} extended to multi-aspect comparison:
\begin{align}
\label{eq:bpmr}
    &\prod\limits_{u\in P}p(>_u|U,V,W,\bm\Sigma)
    = \prod\limits_{(u,i,j)\in P\times Q\times Q}p(\bm D_{uij}\cdot D_{uij}>0|U,V,W,\bm\Sigma) \\\nonumber
    &= \prod\limits_{(u,i,j)\in P\times Q\times Q}\int_0^\infty p(\bm D_{uij}\cdot D_{uij}|U,V,W,\bm\Sigma)d\bm D_{uij}\cdot D_{uij} \\\nonumber
    &= \prod\limits_{(u,i,j)\in P\times Q\times Q}\int_0^\infty \mathcal{N}(\bm D_{uij}\cdot D_{uij}|(U_u * (V_i-V_j))W^T \cdot D_{uij}, D_{uij}\bm\Sigma_{uij}D_{uij}^T)d\bm D_{uij}\cdot D_{uij} \\\nonumber
    &= \prod\limits_{(u,i,j)\in P\times Q\times Q}\frac{1}{2}\Big[1 - erf(\frac{- (U_u * (V_i-V_j))W^T \cdot D_{uij}}{\sqrt{2}\sqrt{D_{uij}\bm\Sigma_{uij} D_{uij}^T}})\Big]
\end{align}
where $erf(x)=\frac{2}{\sqrt\pi}\int_{0}^{x}e^{-t^2}dt$ is the Gauss error function. Though the error function could not be expressed as elementary functions, the derivative of it can be explicitly calculated by $\frac{d}{dx}erf(x)=\frac{2}{\sqrt\pi}e^{-x^2}$. Therefore we could apply gradient based methods to optimize the objective function. 
By treating the latent factors as random variables, we place zero-mean spherical Gaussian priors on the three latent factor matrices $U,V,W$ \cite{PSMF, PMF}:
\begin{equation*}
    p(U|\sigma_U^2) = \prod\limits_{u=1}^M \mathcal{N}(U_u|0,\sigma_U^2 \bm I),~~p(V|\sigma_V^2) = \prod\limits_{v=1}^N \mathcal{N}(V_v|0,\sigma_V^2 \bm I),~~p(W|\sigma_W^2) = \prod\limits_{k=1}^K \mathcal{N}(W_k|0,\sigma_W^2 \bm I) 
\end{equation*}
We also impose the normal-inverse-Wishart distribution which is defined on real-valued positive (semi-)definite matrices as the prior for the covariance matrices \cite{Conjugate,wishart}:
\begin{equation*}
    p(\bm\Sigma|\Psi,\nu) = \frac{|\Psi|^{\nu/2}}{2^{\nu K/2}\Gamma_K(\frac{\nu}{2})}|\bm \Sigma|^{-(\nu + K + 1)/2}e^{\frac{1}{2}tr(\Psi\bm\Sigma^{-1})}
\end{equation*}
where $|X|$ is the determinant of matrix $X$ and $tr(X)$ is the trace of $X$. $\nu$ and $\Psi$ are hyper-parameters which denote that the prior covariance matrix is estimated from $\nu$ observations and with sum of pairwise deviation products $\Psi$. Then the logarithm of the posterior distribution over all latent factors and the covariance matrices is:
\begin{align}
\label{eq:bpmr-personalized}
    \mbox{ln}\,p(&U,V,W,\bm\Sigma|R,\nu_p,\Psi_p,\sigma_U^2,\sigma_V^2,\sigma_W^2)\\ \nonumber
    = &\sum\limits_{(u,i,j)\in P\times Q\times Q} \mbox{ln}\Big[1 - erf(\frac{-(U_u * (V_i-V_j))W^T \cdot D_{uij}}{\sqrt{2}\sqrt{D_{uij}\bm\Sigma_{uij} D_{uij}^T}})\Big] \\ \nonumber
    & +\sum\limits_{i=1}^M \big[-\frac{\nu_p+K+1}{2}\mbox{ln}|\bm\Sigma^U_i| + \frac{1}{2}tr(\Psi_p{\bm\Sigma_i^{U}}^{-1}) + \frac{\nu_p}{2}\mbox{ln}|\Psi_p| - \mbox{ln}\,2^{\nu_p K/2}\Gamma_K(\frac{\nu_p}{2})\big]\\\nonumber 
    & +\sum\limits_{j=1}^N \big[-\frac{\nu_p+K+1}{2}\mbox{ln}|\bm\Sigma^V_j| + \frac{1}{2}tr(\Psi_p{\bm\Sigma_j^{V}}^{-1}) + \frac{\nu_p}{2}\mbox{ln}|\Psi_p| - \mbox{ln}\,2^{\nu_p K/2}\Gamma_K(\frac{\nu_p}{2})\big]\\\nonumber 
    & + \mbox{ln}\,p(U|\sigma_U^2) + \mbox{ln}\,p(V|\sigma_V^2) + \mbox{ln}\,p(W|\sigma_W^2) + C
\end{align}
where $\bm\Sigma$ is the set of covariance matrices for all users and items. $\mbox{ln}\,p(U|\sigma_U^2) = -\frac{1}{2}(\frac{1}{\sigma_U^2}\sum\limits_{u=1}^MU_uU_u^T + Md\,\mbox{ln}\sigma_U^2)$, and it is similar for $V$ and $W$. $C$ is a constant that does not depend on the latent factors. Nevertheless, the covariance matrices need sufficient observations to estimate. For the users and items with only a few observations, it is intractable to estimate accurate covariance matrices. To address the problem of lacking enough observations for estimating the latent factors for such users and items, we introduce an adaptive global covariance matrix $\bm\Sigma_G$ estimated across all users and items to set prior matrix $\Psi_p$. The global covariance matrix can be obtained by optimizing:
\begin{align}
\label{eq:bpmr-global}
    \mbox{ln}\,p(&U,V,W,\bm\Sigma_G|R,\nu_g,\Psi_g,\sigma_U^2,\sigma_V^2,\sigma_W^2)\\ \nonumber
    = &\sum\limits_{(u,i,j)\in P\times Q\times Q} \mbox{ln}\Big[1 - erf(\frac{- (U_u * (V_i-V_j))W^T \cdot D_{uij}}{\sqrt{2}\sqrt{D_{uij}\bm\Sigma_{G} D_{uij}^T}})\Big] \\ \nonumber
    & -\frac{\nu_g +K+1}{2}\mbox{ln}|\bm\Sigma_G| + \frac{1}{2}tr(\Psi_g\bm\Sigma_G^{-1}) + \frac{\nu_g}{2}\mbox{ln}|\Psi_g| - \mbox{ln}\,2^{\nu_g K/2}\Gamma_K(\frac{\nu_g}{2})\\ \nonumber
    & + \mbox{ln}\,p(U|\sigma_U^2) + \mbox{ln}\,p(V|\sigma_V^2) + \mbox{ln}\,p(W|\sigma_W^2) + C
\end{align}
Then the hyper-parameter $\Psi_p$ for Eq. \eqref{eq:bpmr-personalized} can be set by $\Psi_p = \nu_p\bm\Sigma_G$. The personalized covariance matrices estimated by Eq. \eqref{eq:bpmr-personalized} can then be viewed as a variant from the global covariance matrix based on their own observations. 
This helps address the sparsity of observations in individual users and items. Maximizing the log-likelihood in Eq. \eqref{eq:bpmr-personalized} with adaptive global covariance matrix as prior for the personalized covariance matrices leads to the final formulation of BPMR.

\paragraph{Information-Theoretic Interpretation} One important benefit of explicitly modeling the dependency among aspect ratings is to enhance the explainability of making recommendations, where correlations are indicators of which aspects (explanatory variables) to select for explaining the overall preference prediction (response variable). Chen et al. \cite{Chen:2017:KFS:3295222.3295438} propose to minimize the trace of the conditional covariance operator and select features that maximally account for the dependence of the response on the covariates. The idea is similar to our strategy of selecting aspects with the highest correlations with overall preference for explanation. Chen et al. \cite{l2x} interpret the explainability from an information-theoretic perspective: the mutual information between the response variable and the explanations should be maximized. From the first term of Eq. \eqref{eq:bpmr-personalized} or \eqref{eq:bpmr-global}, we can separate out a term $\mbox{ln} |\bm\Sigma|$ in the error function. It turns out that it is consistent with such information-theoretic interpretations, which provides a lower bound to the mutual information among all aspects. Specifically, we have:
\begin{theorem}
\label{theorem}
Given the variance vector $\bm\sigma = (\sigma_{x1}, \sigma_{x2}, \dots , \sigma_{xK})^T$ of $K$ random variables, the mutual information of $K$ multivariate variables $\bm X\in\mathbb R^K$ with covariance matrix $\bm\Sigma$ is lower bounded by $-\mbox{ln}|\bm\Sigma|$:
\begin{equation}
    I(\bm X) \geq -\frac{1}{2}\mbox{ln}|\bm\Sigma| + C
\end{equation}
with the equality holds when only one single variable is correlated with the other $K-1$ variables while the correlations among all the other $K-1$ variables are $0$. 
\end{theorem}
The proof is provided in Appendix A. Theorem \ref{theorem} means that maximizing the likelihood of observed orders under the PMTF introduces an information-theoretic regularization term that indirectly maximizes the mutual information among different aspects. Especially, when the correlation only happens between the overall preference and aspect preference, while all aspect preferences are independent, the term of $-\mbox{ln}|\bm\Sigma|$ is exactly equivalent to the mutual information. 
When maximizing the likelihood defined in Eq. {eq:bpmr-personalized}, we explicitly maximize $-\mbox{ln}|\bm\Sigma|$, and therefore optimizing the mutual information from below. 
Such information-theoretic interpretation confirmed the rationale of our design and guarantees the explainability of ranking aspects based on PMTF. 

\subsection{Optimization}
We adopt an Expectation-Maximization algorithm to estimate the latent factor matrices and the covariance matrices simultaneously. In E-step, we perform stochastic gradient descent (SGD) to update the latent factor matrices $U,V,W$. In M-step, we first update the global covariance matrix with SGD based on Eq. \eqref{eq:bpmr-global}, and then use the global covariance matrix as a prior parameter for updating the personalized covariance matrices based on Eq. \eqref{eq:bpmr-personalized}. The time complexity of training scales linearly with the number of users, items, aspects, and dimension of latent factors. Notice that in order to learn valid covariance matrices, we need to ensure that they are symmetric and positive semi-definite \cite{covariance}. To satisfy this constraint, we set $\bm\Sigma = LL^T$, where L is an arbitrary matrix of the same dimensionality $\mathbb R^{K\times K}$ with $\bm\Sigma$. Then for any non-zero column vector $\bm x$, we have $\bm x^TLL^T\bm x = (L^T\bm x)^T(L^T\bm x) \geq 0$, which ensures the positive semi-definite property of $LL^T$. And the symmetric property is naturally satisfied as $LL^T = (LL^T)^T$. With this decomposition of covariance matrix, we update the matrix $L$ in the optimization procedure and construct the covariance matrices with the resulting $L$. Derivatives with respect to all parameters based on Eq. \eqref{eq:bpmr-global} and \eqref{eq:bpmr-personalized} are provided in Appendix B.

To optimize for a ranking task, we have in total $O(M\times N\times N)$ training triples of $(u,i,j)$, which make it necessary to find an efficient sampling strategy for SGD. Follow the solution in \cite{BPR}, we adopt bootstrap sampling with replacement of training triples, which samples triples in each iteration uniformly at random. As two unobserved items of a user cannot determine a preference order, such pairs can be removed from model training. We terminate the EM iterations based on the change of objective function on a hold-out validation data set. 

%% file: exp.tex
\section{Experiment Results}
\label{sec:exp}
In this section, we evaluate our proposed solution on a large multi-aspect rating dataset collected from TripAdvisor reviews. We compare our general PMFT for prediction task and BPMF for ranking task with several existing models. Besides standard metrics on rating prediction and item ranking, we also propose a metric on the consistency of explanations composed with selected aspects. Both quantitative and qualitative results show the effectiveness and advantage of our proposed method. 
\subsection{TripAdvisor Multi-aspect Rating Dataset}
The data was collected from May 2014 to September 2014 from TripAdvisor\footnote{http://www.tripadvisor.com Data was collected before they forbid practitioners to crawl their data.}. There are over 3 million multi-aspect ratings from over 1 million users and around 18 thousand hotels. Each multi-aspect rating vector consists of one overall rating and eight detailed aspect ratings: [`\textit{Sleep Quality}', `\textit{Service}', `\textit{Value}', `\textit{Rooms}', `\textit{Cleanliness}', `\textit{Location}', `\textit{Check in / front desk}', `\textit{Business service}']. All ratings range from 1 to 5. After filtering out users and hotels who have less than 5 rating vectors, we obtain the final dataset of 1,057,217 multi-aspect ratings from 100,005 users and 17,257 hotels. We split the dataset by using 70\% for training, 15\% for validation and 15\% for testing. The processed data will be made available online. 

\subsection{Baselines \& Experiment Settings}
We selected several popular baselines that model multi-aspect ratings:
\begin{itemize}
    \item[-] PTF: Probabilistic Tensor Factorization \cite{bptf,Rai:2015:SPT:2832747.2832775}. Traditional tensor factorization based on CP decomposition, which treats all aspects independently.
    \item[-] BPR: Bayesian Personalized Ranking \cite{BPR}. A generic ranking criterion that maps the difference between predicted ratings by a sigmoid function. For multi-aspect ranking, we apply BPR on each aspect separately.
    \item[-] EFM: Explicit Factor Models \cite{EFM}. It is a joint matrix factorization algorithm that models user's aspect attention together with item's aspect quality.
    \item[-] MTER: Explainable Recommendation via Multi-task Learning \cite{MTER}. A joint tensor factorization algorithm that models user's detailed opinions on individual aspects. As in this work we does not consider users' detailed review content, we exclude its content modeling component and apply BPR to the overall ratings.
\end{itemize}
The implementations of all baseline models and our proposed model will all be available online.

After exploring different combinations of model hyper-parameters, we set the latent factor dimension to $d=13$, $\sigma_U =\sigma_V =\sigma_W = 1$ and $\lambda = 0.5$. For the inverse-Wishart priors, we set $\nu_g = 50,000$ for global covariance and $\nu_p = 10$ for personalized covariances. To impose the prior for the global covariance matrix, we first estimate the covariance matrix $\bm\Sigma_g$ from all observed multi-aspect ratings across users and items, and set $\Psi_g = \nu_g\bm\Sigma_g$. Each EM step consists of 5 iterations of SGD update, with 1,000 samples of pair-wise comparisons per iteration. For better convergence, we also apply AdaGrad \cite{adagrad} with an initial learning rate of 0.03. 

\subsection{Results on Top-K Recommendation}
To compare the ranking performance, we adopt Normalized Discounted Cumulative Gain (NDCG) as the metric. We set the candidate item list for each user to 150 items and report average NDCG across all users. The candidate list is composed of all observed items in the testing set and randomly selected items for the rest. All the randomly selected items will be treated as irrelevant in evaluation. We evaluate Top-K ranking performance at 10, 20 and 50. Due to the space limit, we only report the ranking performance based on overall preference, and the results on all aspects are reported in Appendix C. Table \ref{tab:ndcg} demonstrates the NDCG results for all baseline models. Three models MTER, BPR and BPMR that directly optimize for ranking generally outperformed PTF, which only optimizes for pointwise rating prediction. Our proposed solution BPMR achieves the highest NDCG on all top-K evaluations with improvements over 5\% compared to the second best model. This improvements from BPMR shows the effectiveness of modeling explicit dependency among aspects and directly maximizing the probability of observed orders between vectorized comparisons.   

\begin{table}[]
\centering
\caption{Ranking performance on NDCG@K. The improvement is from the best v.s. the second best.}
\label{tab:ndcg}
\begin{tabular}{@{}ccccccc@{}}
\toprule
NDCG & EFM    & MTER   & PTF    & BPR    & BPMR            & Improvement \\ \midrule
@10  & 0.1579 & 0.1874 & 0.1067 & 0.1981 & \textbf{0.2116} & 6.81\%*     \\ \midrule
@20  & 0.1802 & 0.2259 & 0.1397 & 0.2425 & \textbf{0.2592} & 6.89\%*     \\ \midrule
@50  & 0.2437 & 0.2884 & 0.2068 & 0.3084 & \textbf{0.3267} & 5.93\%*     \\ \bottomrule
\end{tabular}
\\ $*p<0.05$
\end{table}

\begin{figure}
    \centering
    \includegraphics[width=2.7in]{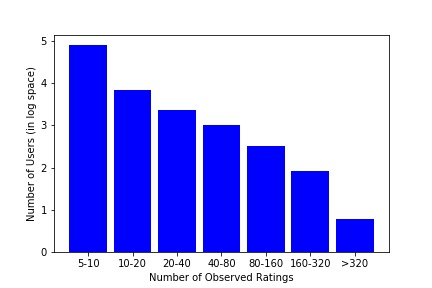}
    \includegraphics[width=2.7in]{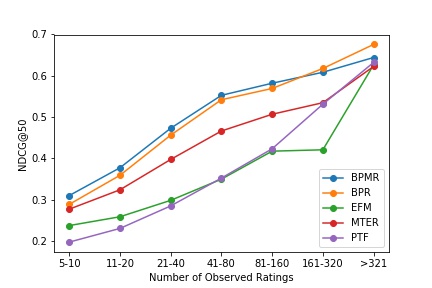}
    \caption{Left: Distribution of users. Right: Performance of models on NDCG@50. Users are grouped by the number of observed ratings in the training data.}
    \label{fig:ndcg_hlusers}
\end{figure}

In order to better understand the ranking performance on different users, we also group users according to the number of observed ratings. Figure \ref{fig:ndcg_hlusers} shows the distribution of grouped users with the x-axis being the number of observed rating vectors and y-axis being the number of users in the group in a log space. The user distribution is highly skewed, as most users (around 80,000 users) have less than 10 observations. Figure \ref{fig:ndcg_hlusers} shows the corresponding performance of NDCG@50 on different user groups. BPMR generally performs better than all other baselines, especially on the light users. The reason is that the explicit introduction of aspect correlations in BPMR enables the learnt user factors in light users to fully exploit the observed aspect ratings to enhance the ranking performance. An interesting observation is than for heavy users with more than 160 observations, the advantage of BPMR is diminishing. One possible reason is that in heavy users there are sufficient observations for baselines to fit all the patterns including the dependencies existing among aspects, even the baselines do not explicitly model such aspect-level dependency. 

\subsection{Learned Correlations for Explainability}
Besides the ranking evaluation, we also propose a consistency metric called Mean Error Correlation (MEC). It measures the consistency between the overall rating and selected aspect rating for explanation. The idea is to calculate the correlations between the prediction error of overall ratings and that of the selected aspect ratings across all user-item pairs in the testing set $T$: 
\begin{equation}
    MEC = \frac{\sum\limits_{(u,i)\in T} (E_{ui} - \overline E)\times(S_{ui}-\overline{S})}{\sqrt{\sum\limits_{(u,i)\in T} (E_{ui} - \overline E)^2}\sqrt{\sum\limits_{(u,i)\in T} (S_{ui} - \overline S)^2}}
\end{equation}
where $E_{ui}=\hat r^{over}_{ui} - r^{over}_{ui}$ is the error of the predicted overall rating and $S_{ui}=\hat r^{sasp}_{ui} - r^{sasp}_{ui}$ is the error of the prediction for a selected aspect. $E$ and $S$ are their means. Such correlation measures the consistency between the predictions of overall preference and the selected aspect preference, and thus reflects to what extent the selected aspect can explain the resulting overall preference. For PTF and BPR, as there is no criterion to select aspects, we randomly select an aspect to explain the recommendation. In EFM, the aspect is selected by the linear combination of user's attention and item's quality. MTER proposes to use the aspect with highest predicted rating as explanation. For BPMR, we use the aspect with the highest corerlation with the overall rating as explanation. Table \ref{tab:mec} reports the MEC result of all models. It is clear that BPMR significantly outperforms all other models on this consistency metric in selecting aspects for explanation. The results proves the validity of the learned correlations. To better understand the learned correlations, we also put a visualization of the learned global covariance matrix in Appendix C, which reflects the general focuses of users and items in the dataset. The average user and item correlation matrices are similar.

\begin{table}[]
\centering
\caption{MEC of overall rating and the selected aspect rating for explanations. }
\label{tab:mec}
\begin{tabular}{@{}ccccc@{}}
\toprule
EFM    & MTER   & PTF    & BPR    & BPMR   \\ \midrule
0.2636 & 0.2963 & 0.2043 & 0.2546 & \textbf{0.4969} \\ \bottomrule
\end{tabular}
\end{table}

%% file: conclusion.tex
\section{Conclusion}
We formulate a multi-aspect ranking criterion that ranks multi-aspect preferences collectively. To maintain the dependence among different aspects, we suggest to use vectorized representations of multi-aspect ratings and propose a probabilistic multivariate tensor factorization framework PMTF. The framework naturally leads to the derivation of Bayesian Probabilistic Multivariate Ranking criterion, which generalizes the single-aspect ranking to a multivariate fashion. 

In practice it is easier to extract users' implicit preference, e.g., click on one item but not another. In this case, it is meaningful to extend our current solution to handle such implicit comparison between items, for example exploring multiple instance learning \cite{maron1998framework}: an item is preferred over another if at least one of its aspects is preferred.  

%% file: appendix.tex
\section*{Appendix A}
\subsection*{The proof of Theorem 1: }
\begin{proof}
The covariance matrix could be written as $\bm\Sigma = (\bm\sigma\bm\sigma^T)*\bm\rho$, where $\bm\rho$ is the correlation matrix constructed from $\bm\Sigma$. With Hadamard product, we have:
\begin{equation*}
    det(\bm\Sigma) = det((\bm\sigma\bm\sigma^T)*\bm\rho) \geq det(\bm\sigma\bm\sigma^T)\cdot det(\bm\rho)
\end{equation*}
As a result, we have,
\begin{equation*}
    \mbox{ln}|\bm\Sigma| \geq \mbox{ln}|\bm\sigma\bm\sigma^T| + \mbox{ln}|\bm\rho|
\end{equation*}
It is proved that $I(\bm X) = -\frac{1}{2}\mbox{ln}|\bm\rho|$ in \cite{mutualinfo}, which leads to,
\begin{equation*}
    \mbox{ln}|\bm\Sigma| \geq \mbox{ln}|\bm\sigma\bm\sigma^T| - 2I(\bm X)
\end{equation*}
Therefore, we can obtain the following,
\begin{equation*}
   I(\bm X) \geq -\frac{1}{2}\mbox{ln}|\bm\Sigma| + C
\end{equation*}

\textit{Proof of equality:}
Assume only 
Denote the first random variable with standard deviation $\sigma_y^2$ as $r_y$, and assume only it is correlated with the other $K-1$ random variables. We denote such correlation as $\bm{c} = (c_{y1}, \dots,c_{y(K-1)})$. The other $K-1$ random variables with standard deviation $\bm\sigma = (\sigma_1,\dots,\sigma_{K-1})$ are denoted as $(r_{1},\dots,r_{K-1})$; and the correlation among themselves is assumed to be zero. Then the covariance matrix can be written as,
\begin{equation*}
    \bm\Sigma = 
    \begin{bmatrix}
    \sigma_y^2               &c_{y1}\sigma_y\sigma_1   &c_{y2}\sigma_y\sigma_2   &\dots &c_{y(K-1)}\sigma_y\sigma_{K-1} \\
    c_{y1}\sigma_y\sigma_1   &\sigma_1^2               &                         &      &        \\
    c_{y2}\sigma_y\sigma_2   &                         &\sigma_2^2               &      &\text{\huge0}       \\
    \vdots                   &                         &                         &\ddots       & \\
    c_{y(K-1)}\sigma_y\sigma_{K-1} &                   &\text{\huge0}                         &      &\sigma_{K-1}^2    \\
    \end{bmatrix}
\end{equation*}

We can calculate the determinant of it by,
\begin{equation*}
    |\bm\Sigma| = 
    \begin{vmatrix}
    \sigma_y^2 (1 - \sum\limits_{i=1}^{K-1} c_{yi}^2)               &c_{y1}\sigma_y\sigma_1   &c_{y2}\sigma_y\sigma_2   &\dots &c_{y(K-1)}\sigma_y\sigma_{K-1} \\
                             &\sigma_1^2               &                         &      &        \\
                             &                         &\sigma_2^2               &      &\text{\huge0}       \\
                             &\text{\huge0}            &                         &\ddots       & \\
                             &                         &                         &      &\sigma_{K-1}^2    \\
    \end{vmatrix} 
   = \sigma_y^2 (1 - \sum\limits_{i=1}^{K-1} c_{yi}^2) \prod\limits_{i=1}^{K-1}\sigma_{i}^2 
\end{equation*}
As a result, we can derive that,
\begin{equation*}
    -\mbox{ln}|\Sigma| = -\mbox{ln}\,\sigma_y^2 - \sum\limits_{i=1}^{K-1}\sigma_i^2 - \mbox{ln}(1 - \sum\limits_{i=1}^{K-1} c_{yi}^2) 
                       = C - \mbox{ln}|\bm \rho|
                       = C + 2I(X)
\end{equation*}
This concludes our proof. 
\end{proof} 

\section*{Appendix B}
\subsection*{Gradients with respect to the model parameters}
We use $\mathcal L(U,V,W,\bm\Sigma)$ to represent the log likelihood of Eq. \eqref{eq:bpmr-global} with global covariance matrix, and the log likelihood of Eq. \eqref{eq:bpmr-personalized} with personalized covariance matrices. $\underset{U,V,W,\bm\Sigma}{\mbox{max}}\mathcal L(U,V,W,\bm\Sigma)$ corresponds to the objective functions. 

\paragraph{Gradients for Eq. \eqref{eq:bpmr-global}:}
Denote $z = \frac{- (U_u * (V_i-V_j))W^T \cdot D_{uij}}{\sqrt{2}\sqrt{D_{uij}\bm\Sigma_{uij} D_{uij}^T}})$, we have the following gradients for model update, 
\begin{equation*}
\begin{split}
    \frac{\partial \mathcal L}{\partial U_u} &= \sum\limits_{(i,j)\in Q\times Q}\frac{2e^{-z^2}}{\sqrt{2\pi}(1-erf(z))}\frac{D_{uij}}{\sqrt{D_{uij}\bm\Sigma_GD_{uij}^T}}((V_i-V_j)*W) - \frac{1}{\sigma_U^2}U_u\\
    \frac{\partial \mathcal L}{\partial V_i} &= \sum\limits_{(u,j)\in P\times Q}\frac{2e^{-z^2}}{\sqrt{2\pi}(1-erf(z))}\frac{D_{uij}}{\sqrt{D_{uij}\bm\Sigma_GD_{uij}^T}}(U_u*W) - \frac{1}{\sigma_V^2}V_i\\
    \frac{\partial \mathcal L}{\partial V_j} &= \sum\limits_{(u,i)\in P\times Q}\frac{-2e^{-z^2}}{\sqrt{2\pi}(1-erf(z))}\frac{D_{uij}}{\sqrt{D_{uij}\bm\Sigma_GD_{uij}^T}}(U_u*W) - \frac{1}{\sigma_V^2}V_j\\
    \frac{\partial \mathcal L}{\partial W} &= \sum\limits_{(u,i,j)\in P\times Q\times Q}\frac{2e^{-z^2}}{\sqrt{2\pi}(1-erf(z))}\frac{D_{uij}^T}{\sqrt{D_{uij}\bm\Sigma_GD_{uij}^T}}(U_u*((V_i-V_j))) - \frac{1}{\sigma_W^2}W\\
    \frac{\partial \mathcal L}{\partial\bm\Sigma_G} &= \sum\limits_{(u,i,j)\in P\times Q\times Q}\frac{-2e^{-z^2}}{\sqrt{2\pi}(1-erf(z))}(U_u * (V_i-V_j))W^T\cdot D_{uij}\frac{D_{uij}^TD_{uij}}{(D_{uij}\bm\Sigma_GD_{uij}^T)^{3/2}} \\
    &+ \frac{1}{2}(\bm\Sigma_G^{-1}\Psi\bm\Sigma_G^{-1}) - \frac{\nu + K + 1}{2}\bm\Sigma_G^{-T}\\
    \frac{\partial \mathcal L}{\partial L_G} &= 2\frac{\partial l}{\partial \bm\Sigma_G}L_G
\end{split}
\end{equation*}

\paragraph{Gradients for Eq. \eqref{eq:bpmr-personalized}:}
\begin{equation*}
\begin{split}
    \frac{\partial \mathcal L}{\partial U_u} &= \sum\limits_{(i,j)\in Q\times Q}\frac{2e^{-z^2}}{\sqrt{2\pi}(1-erf(z))}\frac{D_{uij}}{\sqrt{D_{uij}\bm\Sigma_{uij}D_{uij}^T}}((V_i-V_j)*W) - \frac{1}{\sigma_U^2}U_u\\
    \frac{\partial \mathcal L}{\partial V_i} &= \sum\limits_{(u,j)\in P\times Q}\frac{2e^{-z^2}}{\sqrt{2\pi}(1-erf(z))}\frac{D_{uij}}{\sqrt{D_{uij}\bm\Sigma_{uij}D_{uij}^T}}(U_u*W) - \frac{1}{\sigma_V^2}V_i\\
    \frac{\partial \mathcal L}{\partial V_j} &= \sum\limits_{(u,i)\in P\times Q}\frac{-2e^{-z^2}}{\sqrt{2\pi}(1-erf(z))}\frac{D_{uij}}{\sqrt{D_{uij}\bm\Sigma_{uij}D_{uij}^T}}(U_u*W) - \frac{1}{\sigma_V^2}V_j\\
    \frac{\partial \mathcal L}{\partial W} &= \sum\limits_{(u,i,j)\in P\times Q\times Q}\frac{2e^{-z^2}}{\sqrt{2\pi}(1-erf(z))}\frac{D_{uij}^T}{\sqrt{D_{uij}\bm\Sigma_{uij}D_{uij}^T}}(U_u*((V_i-V_j))) - \frac{1}{\sigma_W^2}W\\
    \frac{\partial \mathcal L}{\partial\bm\Sigma_{uij}} &= \sum\limits_{(u,i,j)\in P\times Q\times Q}\frac{-2e^{-z^2}}{\sqrt{2\pi}(1-erf(z))}(U_u * (V_i-V_j))W^T\cdot D_{uij}\frac{D_{uij}^TD_{uij}}{(D_{uij}\bm\Sigma_{uij}D_{uij}^T)^{3/2}} \\
    \frac{\partial \mathcal L}{\partial\bm\Sigma_{u}^U} &= 2\lambda \frac{\partial \mathcal L}{\partial\bm\Sigma_{uij}} + \frac{1}{2}({\bm\Sigma_{u}^U}^{-1}\Psi{\bm\Sigma_{u}^U}^{-1})^T - \frac{\nu+K+1}{2}{\bm\Sigma_{u}^U}^{-T}\\
    \frac{\partial \mathcal L}{\partial\bm\Sigma_{i}^V} &= (1-\lambda) \frac{\partial \mathcal L}{\partial\bm\Sigma_{uij}} + \frac{1}{2}({\bm\Sigma_{i}^V}^{-1}\Psi{\bm\Sigma_{i}^V}^{-1})^T - \frac{\nu+K+1}{2}{\bm\Sigma_{i}^V}^{-T} \\
    \frac{\partial \mathcal L}{\partial\bm\Sigma_{j}^V} &= (1-\lambda) \frac{\partial \mathcal L}{\partial\bm\Sigma_{uij}} + \frac{1}{2}({\bm\Sigma_{j}^V}^{-1}\Psi{\bm\Sigma_{j}^V}^{-1})^T - \frac{\nu+K+1}{2}{\bm\Sigma_{j}^V}^{-T}\\
    \frac{\partial \mathcal L}{\partial L_u^U} &= 2\frac{\partial l}{\partial \bm\Sigma_u^U}L_u^U,\,\,\,\,\frac{\partial \mathcal L}{\partial L_i^V} = 2\frac{\partial l}{\partial \bm\Sigma_i^V}L_i^V,\,\,\,\,\frac{\partial \mathcal L}{\partial L_j^V} = 2\frac{\partial l}{\partial \bm\Sigma_j^V}L_j^V
\end{split}
\end{equation*}
where $\Psi = \nu\bm\Sigma_G$ and $\bm\Sigma_{uij} = 2\lambda\bm\Sigma_{u}^U+(1-\lambda)(\bm\Sigma_{i}^V+\bm\Sigma_{j}^V)$.

\section*{Appendix C}
\subsection*{Visualization of the learned correlation matrix}
\begin{figure}
    \centering
    \includegraphics[width = 4in]{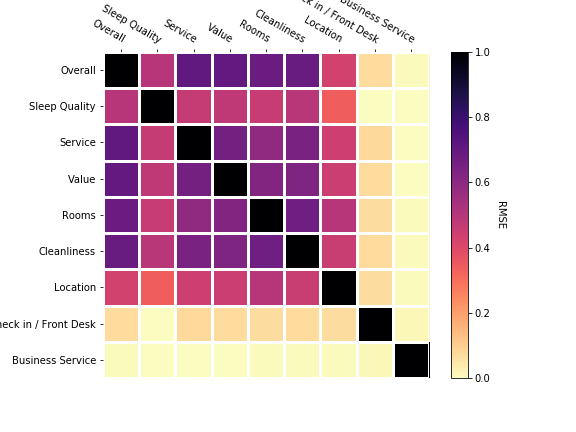}
    \caption{Learned Global Correlations}
    \label{fig:cov}
\end{figure}
Figure \ref{fig:cov} shows the learned global correlation matrix from BPMR. From the figure we can tell that the aspects [`\textit{Service}', `\textit{Value}', `\textit{Rooms}', `\textit{Cleanliness}'] are highly correlated with the overall preference, while [`\textit{Check in / front desk}', `\textit{Business service}'] are the least correlated aspects. This results is quite consistent with our intuition, for most people will care about the aspects like `Rooms' and `Cleanliness' when evaluating the experience of living in a hotel, but in general `Check in / front desk', `Business service' will not affect our experience that much. This qualitative result of learned correlations further validates the effectiveness of introducing correlations and learning it with BPMR.

\subsection*{Ranking performance on all aspects}
\begin{figure}
    \centering
    \includegraphics[width = 1.8in]{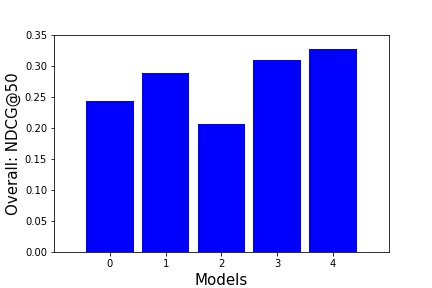}
    \includegraphics[width = 1.8in]{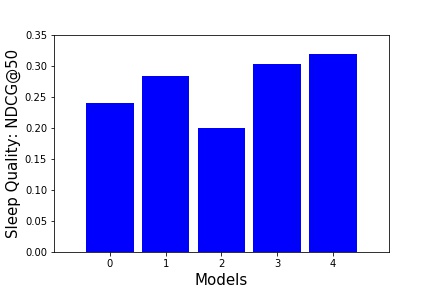}
    \includegraphics[width = 1.8in]{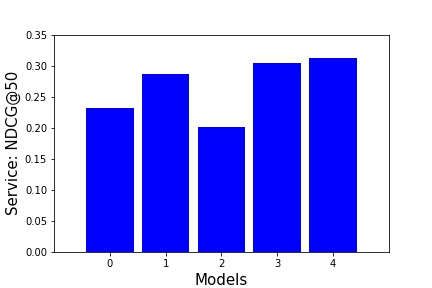}
    \includegraphics[width = 1.8in]{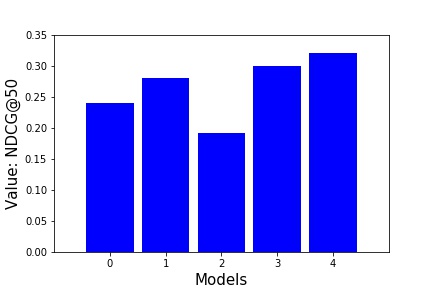}
    \includegraphics[width = 1.8in]{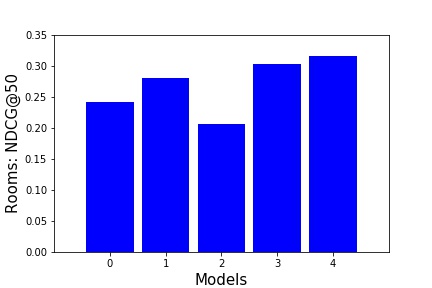}
    \includegraphics[width = 1.8in]{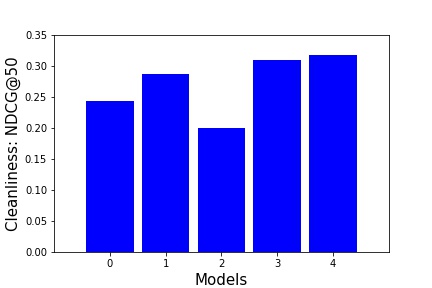}
    \includegraphics[width = 1.8in]{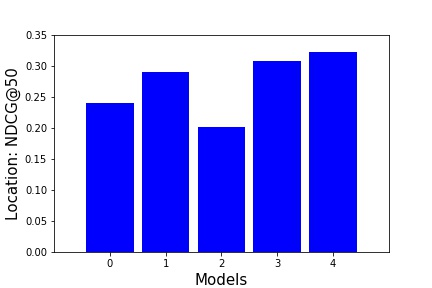}
    \includegraphics[width = 1.8in]{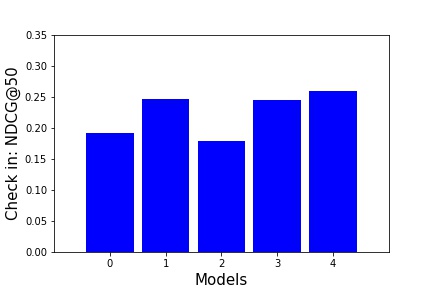}
    \includegraphics[width = 1.8in]{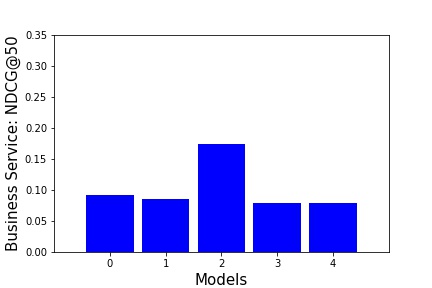}
    \caption{Ranking performance of all aspects.}
    \label{fig:ndcg}
\end{figure}

We have one overall rating and eight aspects: [`\textit{Sleep Quality}', `\textit{Service}', `\textit{Value}', `\textit{Rooms}', `\textit{Cleanliness}', `\textit{Location}', `\textit{Check in / front desk}', `\textit{Business service}']. The ranking performance of each aspect on NDCG@50 is shown in Figure \ref{fig:ndcg}. Models are represented as: [0: EFM, 1: MTER, 2: PTF, 3: BPR, 4:BPMR].
From the results we can see that BPMR outperforms other models on most aspects consistently. However, on the aspects \textit{'Check in/front desk'} and \textit{`Business Service'}, the performances of all models are much worse than other aspects. This might be because the preferences of these two aspects are correlated with others to a very small extent, as suggested in Figure \ref{fig:cov}. Therefore, these two aspects have lower weights introduced by the correlation in the objected function and thus not well optimized. So the results on these two aspects are not consistent with other aspects. 
